\documentclass[fleqn,usenatbib]{mnras}

\usepackage{newtxtext,newtxmath,color}
\usepackage{threeparttable}
\usepackage{multirow}
\usepackage{booktabs}
\usepackage{ulem}
\usepackage{makecell}

\usepackage[T1]{fontenc}

\DeclareRobustCommand{\VAN}[3]{#2}
\let\VANthebibliography\thebibliography
\def\thebibliography{\DeclareRobustCommand{\VAN}[3]{##3}\VANthebibliography}

\usepackage{graphicx}	% Including figure files
\usepackage{subcaption}
\usepackage{amsmath}	% Advanced maths commands
\usepackage{amssymb}	% Extra maths symbols
\usepackage{cellspace}
\usepackage{makecell}
\usepackage{longtable}
\title[]{CURLING - \uppercase\expandafter{\romannumeral1}. The Influence of Point-like Image Approximation on the Outcomes of Cluster Strong Lens Modeling}

\author[Y.-S. Xie et al.]{
Yushan Xie$^{1, 3}$,
Huanyuan Shan$^{1, 3, 4}$\thanks{E-mail: \url{hyshan@shao.ac.cn}},
Nan Li$^{2}$\thanks{E-mail: \url{nan.li@nao.cas.cn}},
Ran Li$^{2, 10, 3}$,
Eric Jullo$^{5}$,
Chen Su$^{1, 3}$,
\newauthor{Xiaoyue Cao$^{2, 3}$,
Jean-Paul Kneib$^{6}$,
Ana Acebron$^{7, 8, 9}$,
Mengfan He$^{1, 2, 3}$,
Ji Yao$^{1}$,}
\newauthor{
Chunxiang Wang$^{2, 10, 3}$, Jiadong Li$^{11}$ and Yin Li$^{12}$}
\\
$^{1}$Shanghai Astronomical Observatory, Chinese Academy of Sciences, Shanghai 200030, China\\
$^{2}$National Astronomical Observatory, Chinese Academy of Sciences, Beijing 100012, China\\
$^{3}$School of Astronomy and Space Science, University of Chinese Academy of Sciences, Beijing 100049, China\\
$^{4}$Key Laboratory of Radio Astronomy and Technology, Chinese Academy of Sciences, A20 Datun Road, Chaoyang District, Beijing, 100101, P. R. China\\
$^{5}$Aix Marseille Univ., CNRS, CNES, LAM, Marseille, France\\
$^{6}$Laboratory of Astrophysics, École Polytechnique Fédérale de Lausanne (EPFL), Observatoire de Sauverny, CH-1290 Versoix, Switzerland\\
$^{7}$ Instituto de Física de Cantabria (IFCA), CSIC-Univ. de Cantabria, Avda. los Castros, s/n, E-39005 Santander, Spain\\
$^{8}$ Departamento de Física Moderna, Universidad de Cantabria, Avda. de los Castros s/n, 39005 Santander, Spain\\
$^{9}$Dipartimento di Fisica, Università degli Studi di Milano, Via Celoria 16, I-20133 Milano, Italy\\
$^{10}$Institute for Frontiers in Astronomy and Astrophysics, Beijing Normal University, Beijing 102206, China\\
$^{11}$Max-Planck-Institut f{\"u}r Astronomie, K{\"o}nigstuhl 17, D-69117 Heidelberg, Germany\\
$^{12}$Department of Mathematics and Theory, Peng Cheng Laboratory, Shenzhen, Guangdong 518066, China
}

\date{Accepted XXX. Received YYY; in original form ZZZ}

\pubyear{2024}

\begin{document}
\label{firstpage}
\pagerange{\pageref{firstpage}--\pageref{lastpage}}
\bibliographystyle{mnras}
\maketitle

% Abstract of the paper
\begin{abstract}
Cluster-scale strong lensing is a powerful tool for exploring the properties of dark matter and constraining cosmological models. However, due to the complex parameter space, pixelized strong lens modeling in galaxy clusters is computationally expensive, leading to the point-source approximation of strongly lensed extended images, potentially introducing systematic biases. Herein, as the first paper of the ClUsteR strong Lens modelIng for the Next-Generation observations (CURLING) program, we use lensing ray-tracing simulations to quantify the biases and uncertainties arising from the point-like image approximation for JWST-like observations. Our results indicate that the approximation works well for reconstructing the total cluster mass distribution, but can bias the magnification measurements near critical curves and the constraints on the cosmological parameters, the total matter density of the Universe $\Omega_{\rm m}$, and dark energy equation of state parameter $w$. To mitigate the biases, we propose incorporating the extended surface brightness distribution of lensed sources into the modeling. This approach reduces the bias in magnification from 46.2 per cent to 0.09 per cent for $\mu \sim 1000$. Furthermore, the median values of cosmological parameters align more closely with the fiducial model. In addition to the improved accuracy, we also demonstrate that the constraining power can be substantially enhanced. In conclusion, it is necessary to model cluster-scale strong lenses with pixelized multiple images, especially for estimating the intrinsic luminosity of highly magnified sources and accurate cosmography in the era of high-precision observations.

\end{abstract}

% Select between one and six entries from the list of approved keywords.
% Don't make up new ones.

%%%%%%%%%%%%%%%%%%%%%%%%%%%%%%%%%%%%%%%%%%%%%%%%%%

%%%%%%%%%%%%%%%%% BODY OF PAPER %%%%%%%%%%%%%%%%%%

\section{Introduction}

Galaxy clusters, with total masses of $\rm 10^{14}\sim 10^{15} M_{\odot}$, represent the most massive self-gravitationally bound structures in the Universe. As predicted by Einstein's General Relativity (GR), these massive clusters exhibit gravitational lensing effects due to their dense mass distribution. This phenomenon leads to the deflection and strong lensing of light rays from background sources, such as distant galaxies, quasars, and even star candidates, resulting in the formation of multiple images or giant arcs \citep{liang14, yuan12, mahler18, caminha17}. Gravitational lensing by galaxy clusters has evolved into a powerful tool for studying various aspects of astrophysics and cosmology, for instance, providing insights into the origin and evolution of galaxies \citep{welch22, mahler19, treu15, mercurio21}, the characteristics of dark matter \citep{massey15, harvey15, caminha19, jauzac19, bergamini23a, he20}, and enabling precise constraints of cosmological parameters \citep{gilmore09, jullo10, acebron17, caminha21, grillo18, grillo20}.

Accurate lens modeling is crucial for achieving reliable scientific results. However, this task is challenging due to the intricate nature of strong lensing clusters. These clusters, characterized by the substantial lensing cross-section, consist of multiple clumpy dark matter haloes and numerous substructures in a dynamically perturbed state. Over the past few decades, efforts have been made to address various sources of systematic errors associated with lens modeling. For example, errors can arise in parametric methods, where the assumption of analytical models will inevitably deviate from the true mass distribution \citep{acebron17, meneghetti17, raney20}. The presence of large-scale structures along the line-of-sight of the lens system can impact lens modeling \citep{daloisio10, chirivi18}, but is often overlooked in most of the lensing analyses. Given the limitations on accurately measuring the magnitude and velocity dispersion of each cluster galaxy, it is not possible to depict each mass component in a cluster lens, instead galaxy scaling relations are adopted on the basis of mass-to-light relations \citep{tfrelation, fjrelation}, the scatter of these galaxy scaling relations can result in biased modeling outputs \citep{bergamini21, granata22}. The accuracy of lens modeling is also sensitive to the measurements of multiple image redshifts, as demonstrated by \cite{acebron17} who analyzed the difference on cosmological constraints between using spectroscopic and photometric data. 

In addition to the systematic errors as mentioned earlier, another source of error in strong lens modeling arises from the positional error associated with multiple images.
%in the Markov chain Monte Carlo (MCMC) analysis of strong lens modeling. 
During the optimization of lens parameters, accurate likelihood computation involves utilizing a pixelated approach to describe the image light profile and taking into account the signal-to-noise ratio of each pixel \citep{dye05, suyu06, kneib11, nightingale18, birrer18, he23}. However, due to the computational limitations, current algorithms for modeling strong lenses at cluster scale simplify the observed multiple images as point sources, assuming a Gaussian distribution for the associated errors. This assumption relies on certain conditions, such as the compactness of  the source galaxy, and the smoothness of the surface brightness profile. Given that the morphology of a lensed object is typically more complex than a simple Gaussian shape, relying on this simplification in lens modeling can introduce notable biases. These biases can subsequently impact the accuracy of the derived lensing properties.

Observational efforts hold the potential to significantly get rid of the systematic errors and improve the accuracy of current lens modeling. For instance, by incorporating the data from Multi-Unit Spectroscopic Explorer (MUSE) on the Very Large Telescope (VLT) \citep{bacon12} of each member galaxy, it becomes possible to estimate galaxy masses with greater accuracy than by solely relying on scaling-relation models. Such data also enables the spectroscopic confirmation of multiple images, thereby mitigating the risks of misidentifying image families and enhancing the robustness of modeling. Consequently, the errors arising from the modeling process are likely to become more pronounced.

%Leveraging the Multi-Unit Spectroscopic Explorer (MUSE) on the Very Large Telescope (VLT) \citep{bacon12} enables the spectroscopic confirmation of multiple images, thereby mitigating the risks of misidentifying image families and enhancing the robustness of modeling. Furthermore, spectroscopic data from member galaxies provides additional constraints by measuring the stellar kinematics \citep{bergamini19, monna17}, thereby the mass of each galaxy can be estimated more accurately than just adopting the scaling-relation model. On the other hand, with the advent of next-generation telescopes (e.g., {\it Euclid} satellite, \citealt{euclid}; Large Synoptic Survey Telescope, LSST, \citealt{lsst}; James Webb Space Telescope, JWST, \citealt{jwst}; China Survey Space Telescope, CSST, \citealt{csst11, csst18}), we anticipate a substantial increase in the number of cluster strong lensing systems. These observational efforts hold the potential to significantly reduce systematic errors and improve the accuracy and precision of current lens modeling. Consequently, the errors arising from the modeling process are likely to become more pronounced.

To investigate the impact of assuming multiple images as point sources on cluster strong lensing analyses, we employ simulated cluster strong lensing systems to study their model-derived mass profiles, magnification maps, and cosmological parameters. The paper is organized as follows. In Section~\ref{section2}, we provide an overview of the simulated cluster lenses. We briefly introduce the theory of strong lensing analyses, including the generation of mass map and magnification map, the constraint on cosmology, and the modeling procedure in Section~\ref{section3}. Our main results are presented in Section~\ref{section4}. We conclude with a discussion and summary of our work in Section~\ref{section5}. Throughout the paper, we consider a flat $w$CDM cosmology ($\Omega_{\Lambda,0} = 0.7$, $\Omega_{\rm m}=0.3$, $h$ = 0.7) with a constant dark energy equation of state parameter $w = w_{0} = -1$. 
 
% End of introduction

\begin{table*}
\centering
    
    \begin{tabular}{c|c|c|c|c|c|c|c}
    \hline
          Component & $\Delta \alpha$  & $\Delta \delta$  & e & $\theta$ & ${\rm r}_{\rm core}$ & ${\rm r}_{\rm cut}$ & $\sigma$ \\ 
           & [$^{\prime \prime}$] & [$^{\prime \prime}$] & & [deg] & [kpc] & [kpc] & [km/s] \\
 \hline
 \multicolumn{8}{l}{MACS0416-like,  $z = 0.35$}\\
 \multicolumn{8}{l}{$\rm N_{s} = 26$, $z_{\rm s, toy} = \{1.6,\ 3.5,\ 6.1\}$}\\
         \hline
         
         Cluster Halo 1 & -0.22 & 0.06 & 0.81 & 143& 37.4& 988& 615\\
         Cluster Halo 2 & 22.67 & -34.27 & 0.89 & 136& 26.3& 988& 466\\
         Cluster Halo 3 & -32.45 & 8.80 & 0.00 & 0& 34.2& 988& 308\\
         Cluster Halo 4 & 22.80 & -48.15 & 0.76 & 122& 66.7& 988& 707\\ 
         Perturber 1 & 31.96 & -65.55 & 0.00 & 0& 4.9& 274& 77\\ 
         Perturber 2 & 13.34 & 2.62 & 0.60 & -46& 4.9& 65& 106\\ 
         scaling relations & N(gal) = 212 & ${\rm m}^{\rm ref}$ = 17.02 & $\rm r_{\rm core}^{\rm ref}$ = 0.74 kpc & $\rm r_{\rm cut}^{\rm ref}$ = 74 kpc& $\sigma^{\rm ref}$ = 210.00 km/s& & \\
 \hline
 \multicolumn{8}{l}{A2744-like,  $z = 0.4$}\\
 \multicolumn{8}{l}{$\rm N_{s} = 22$, $z_{\rm s, toy} = \{1.5,\ 2.4,\ 3.8\}$}\\
         \hline
         
         Cluster Halo 1 & -1.42 & 0.55 & 0.59 & 91 & 28.6 & 1500 & 515 \\
         Cluster Halo 2 & -17.85 & -15.22 & 0.40 & 54 & 34.1 & 1600 & 632 \\
          Ext. Clump 1 & 99.49 & 85.97 & 0.00 & 0 & 1.3 & 800 & 111 \\
          Ext. Clump 2 & 138.28 & 99.87 & 0.00 & 0 & 1.4 & 800 & 372 \\
          Ext. Clump 3 & 24.23 & 155.84 & 0.00 & 0 & 1.4 & 800 & 294 \\
          BCG-N & 0.0 & 0.0 & 0.28 & 133 & 1.3 & 800 & 208 \\
          BCG-S & -17.95 & -20.05 & 0.74 & 26 & 0.2 & 178 & 308 \\
          scaling relations & N(gal) = 223 & ${\rm m}^{\rm ref}$ = 17.34 & $\rm r_{\rm core}^{\rm ref}$ = 0.15 kpc & $\rm r_{\rm cut}^{\rm ref}$ = 19.52 kpc & $\sigma^{\rm ref}$ = 252.66 km/s& & \\
          \hline
 \multicolumn{8}{l}{MACS1206-like, $z = 0.45$}\\
 \multicolumn{8}{l}{$\rm N_{s} = 25$, $z_{\rm s, toy} = \{1.9,\ 4.2,\ 5.7\}$}\\
         \hline
          Cluster Halo 1 & -1.40 & 0.14 & 0.72 & 19 & 35.2 & 1152 & 748 \\
          Cluster Halo 2 & 9.20 & 3.63 & 0.46 & 117 & 77.6 & 1152 & 663 \\
         Ext. Clump 1 & -28.87 & -6.83 & 0.33 & -25 & 70.9 & 1152 & 501 \\
         scaling relations & N(gal) = 258 & ${\rm m}^{\rm ref}$ = 17.19 & $\rm r_{\rm core}^{\rm ref}$ = 0.86 kpc & $\rm r_{\rm cut}^{\rm ref}$ = 86 kpc & $\sigma^{\rm ref}$ = 210.0 km/s& & \\
          \hline
 \multicolumn{8}{l}{AS1063-like,  $z = 0.5$ }\\
 \multicolumn{8}{l}{$\rm N_{s} = 23$, $z_{\rm s, toy} = \{2.4,\ 3.3,\ 4.3\}$}\\
         \hline
          Cluster Halo 1 & 1.44 & -0.73 & 0.63 & -39 & 111.3 & 1220 & 1165 \\
         Cluster Halo 2 & -48.60 & 26.26 & 0.01 & 0 & 30.5 & 1220 & 213 \\
         Ext. Clump 1 & 18.90 & -73.36 & 0.80 & -162 & 8.6 & 1155 & 356 \\
          BCG & -18.05 & 13.47 & 0.13 & -28 & 221.7 & 2070 & 443 \\
          Perturber & 0.20 & -1.24 & 0.34 & -15 & 88.1 & 610 & 250 \\
          scaling relations & N(gal) = 222 & ${\rm m}^{\rm ref}$ = 16.18 & $\rm r_{\rm core}^{\rm ref}$ = 0.91 kpc & $\rm r_{\rm cut}^{\rm ref}$ = 92 kpc & $\sigma^{\rm ref}$ = 210.0 km/s & & \\
          \hline
         
    \end{tabular}
    \caption{Model parameters for the four cluster lens systems in the simulation. $z$ represents the cluster redshift, $\rm N_{s}$ is the number of lensed sources put in the lens system, and $z_{\rm s, toy}$ denotes the source redshift distribution used in the analyses. Note that in each system, only three of the lensed sources are used for lens modeling due to computational constraints. Within a lensed system, each line provides the parameters for a specific component, where $\Delta \alpha$ and $\Delta \delta$ are the respective relative position to the cluster centre; e is the ellipticity of the clump calculated as e = 1 - $b/a$, $a$ and $b$ are the semi-major and semi-minor axis; $\theta$ is the position angle; $\rm r_{core}$, $\rm r_{cut}$, and $\sigma$ are the core radius, cut radius and velocity dispersion; ${\rm m}^{\rm ref}$, $r_{\rm core}^{\rm ref}$, $\rm r_{\rm cut}^{\rm ref}$, $\sigma^{\rm ref}$ are parameters for computing the scaling relations of member galaxies in Eq.\ref{eq2}.}
    \label{table1}
\end{table*}

\begin{figure}
	% To include a figure from a file named example.*
	% Allowable file formats are eps or ps if compiling using latex
	% or pdf, png, jpg if compiling using pdflatex
	\includegraphics[width=0.49\textwidth]{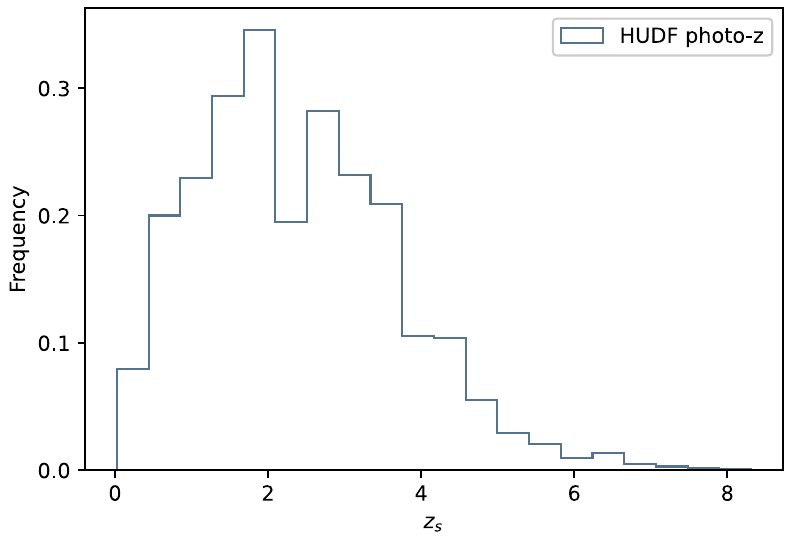}
    \caption{HUDF source redshift distribution. The redshifts of the lensed sources in our sample are assigned according to the HUDF photometric redshift catalog. We exclude the redshifts below $z$ = 1 due to their low lensing efficiency.}
    \label{fig1}
\end{figure}

\section{Lensing ray-tracing simulations} \label{section2}

In this section, we outline the process of setting up the simulation of cluster lens systems. In the sense that stacking the constraints from individual clusters can be an effective approach to yield competitive estimates of the values of the cosmological parameters, we include four lens systems in the simulation, each consisting of two parts: the galaxy cluster acting as the strong lens, and the background sources within the cluster field that are lensed into multiple images.

\subsection{Cluster lenses}\label{subsection21}

To fully exploit the potential of cluster strong gravitational lensing effect, it is crucial to select massive clusters with higher lensing efficiency as the targets. In light of this, we choose to build the clusters based on the strong lensing clusters that have been extensively studied as part of the Hubble Frontier Field (HFF) program \citep{lotz17}. We replicate the publicly available best-fit model parameter sets and the galaxy catalogs\footnote{https://www.fe.infn.it/astro/lensing/} of MACSJ0416.1-2403 \citep{bergamini21}, Abell 2744 \citep{bergamini23a}, MACSJ1206.2-0847, and Abell S1063 \citep{bergamini19}, to describe the mass distribution of the four mock clusters. These referenced models, built with high-quality and extensive spectro-photometric data, along with the largest number of secure multiple images identified so far, provide robust mass distributions for depicting clusters with great strong lensing strengths.

The mass of a mock cluster is thus composed of several components: a) dark matter haloes associated with the large-scale, smooth clumps, b) external clumps representing surrounding substructures within the cluster’s strong lensing field, c) the brightest cluster galaxies (BCGs), and d) galaxy-scale subhaloes corresponding to the member galaxies within the cluster.

Each mass component in the cluster is constructed using an analytic Pseudo Isothermal Elliptical Mass Distribution
 \citep[][PIEMD]{kassiola93, eliasdottir07} profile. The PIEMD density is expressed as: 
\begin{equation}
	\rho(r) = \frac{\rho_{0}}{(1+\frac{r^2}{r_{\rm core}^2})(1+\frac{r^2}{r_{\rm cut}^2})}, 
\end{equation}
where $\rho_{0}$ represents the central density of the profile, $r_{\rm core}$ and $r_{\rm cut}$ are the core radius and the truncation radius, respectively. Within the regions where $r_{\rm core} < r < r_{\rm cut}$, the profile behaves as $\rho \sim r^{-2}$, while it transitions to $\rho \sim r^{-4}$ in the outer regions where $r > r_{\rm cut}$. To define the potential of a PIEMD profile in \textsc{Lenstool} \citep{jullo07}, we utilize a parameter set consisting of \{ $x$, $y$, $e$, $\theta$, $\rm r_{core}$, $\rm r_{cut}$, $\sigma$ \}, where $x$, $y$ describe the centre of the potential, $e$ and $\theta$ define its ellipticity and position angle, $\sigma$ represents the velocity dispersion. The masses of the galaxy-scale subhaloes are determined based on the scaling relations \citep{granata22}, that are related to their respective galaxy magnitudes, following the equations:
\begin{align}\label{eq2}
    \begin{split}
        \left \{
        \begin{array}{ll}
        \rm  \sigma^{gal}_{\it{i}}=\sigma^{ref} 10^{0.4 \frac{m^{ref}-m_{\it{i}}}{\alpha}}\\
        \rm r^{gal}_{core,\it{i}}=r_{core}^{ref}10^{0.4 \frac{m^{ref}-m_{\it{i}}}{2}}\\
        \rm r^{gal}_{cut,\it{i}}=r_{cut}^{ref}10^{0.4 \frac{2(m^{ref}-m_{\it{i}})}{\beta}},
        \end{array}
        \right.
    \end{split}
\end{align}
we fix the parameters of slopes $\alpha = 4.0$ and $\beta = 4.0$.

Given that the typical redshift of a strong lensing cluster is $z \simeq 0.2 - 0.5$ \citep{robertson20}, and the strongest cluster lenses peak at $z \simeq 0.4$ \citep{fox22}, we set the redshifts of the four mock clusters to $z$ = [0.35, 0.4, 0.45, 0.5].

We summarize the properties of the mock clusters in Table~\ref{table1}.

\begin{figure}
	% To include a figure from a file named example.*
	% Allowable file formats are eps or ps if compiling using latex
	% or pdf, png, jpg if compiling using pdflatex
	\includegraphics[width=0.49\textwidth]{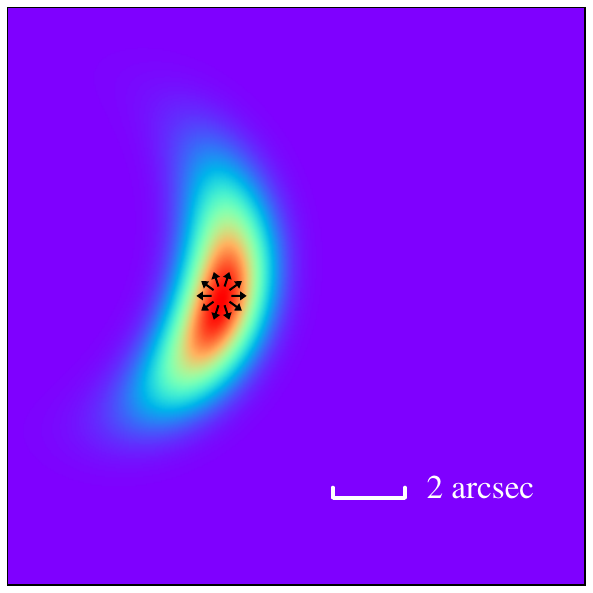}
    \caption{An illustration of the different appearances between the arc-like image (rainbow), generated by the great potential of the intervening lens, and the point-like image (black circle) under the assumption of the observed multiple images being compact point sources.}
    \label{fig2}
\end{figure}

\begin{figure*}
	% To include a figure from a file named example.*
	% Allowable file formats are eps or ps if compiling using latex
	% or pdf, png, jpg if compiling using pdflatex
	\includegraphics[width=0.98\textwidth]{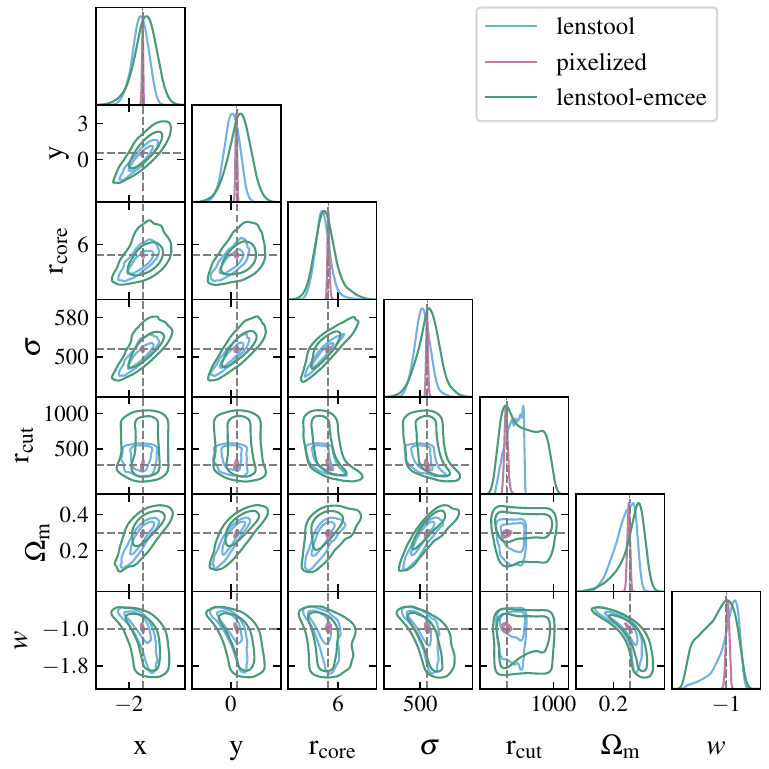}
    \caption{Posteriors obtained with \textsc{Lenstool} (blue), pixelated modeling (pink) and lenstool-emcee (green) for the parameters of cluster main halo and cosmology. The plotted contours are 1 $\sigma$ and 2 $\sigma$ levels.Truths are indicated as the grey dashed lines.}
    \label{fig3}
\end{figure*}

\subsection{Background sources}\label{section22}

In order to simulate the multiple images lensed by the cluster in each system, it is important to not only define the properties of the lens, but also characterize the population of background sources, including their number counts, redshifts, and spatial distribution.

First, to address the number of multiply-lensed sources in a cluster field, we rely on the deeply resolved cluster images obtained from HFF and JWST \citep{bergamini23a, mahler22, zitrin13, caminha22}, which have depths of up to 30 AB magnitude, similar to the expected depths of next-generation surveys for studying strong lensing clusters. These surveys have been demonstrated to have the ability of observing up to hundreds of multiple image families. However, in addition to the identification of multiple images under the consideration of observational depths, we also emphasize the importance of extensive spectroscopic follow-up observations at the cores of galaxy clusters. This is crucial for ensuring robust measurement of the source redshifts, which is essential for accurate lens modeling \citep{grillo15, bergamini21}. Based on these criteria, we assume that a cluster contains more than 20 image families that can be identified and spectroscopically verified.

The source redshift distribution is generated based on the photometric redshift catalog of the Hubble Ultra-Deep Field \citep[HUDF, ][]{beckwith06}, as shown in Figure~\ref{fig1}. Note that the redshift distribution of HUDF galaxies starts from $z = 0.025$, however, in our simulation, we exclude the redshifts below a certain threshold due to their low lensing efficiency. As a result, the sources in our sample fall within the redshift range of $z = 1.0 - 8.4$.

In each lens system, we use the same redshift distribution for the sources, while their positions vary due to the diverse mass distributions of the cluster lenses. The source positions are determined using the ray-tracing algorithm outlined in \cite{li16}, which involves numerically solving the lens equation. With the mass distribution of the lens as introduced in Section~\ref{subsection21} ready, next is to calculate the deflection angle map and build the lens equation. The final step is to map the locations of pixels on the image plane to the source plane, by placing the source on the source plane, light rays of the source are mapped from the source plane onto the image plane again to generate the lensed image. A source position is determined provided that more than one lensed image is generated. The ray-tracing and source-locating are performed with a pixel size of 0.05$^{\prime\prime}$ within a field of $180^{\prime \prime} \times 180 ^{\prime \prime}$. We note that some of the images are missed due to grid resolution in \textsc{Lenstool} while performing lens modeling. Accordingly, we exclude the sources that are found by \textsc{Lenstool} with less than one image from the catalog. Therefore, the number of lensed sources in each lensing system is slightly different, we summarize the number as $\rm N_{s}$ in Table~\ref{table1}.

In addition to determining the positions of the sources, for the novel modeling method proposed in this work, we also compute the extended source surface brightness distribution with the above ray-tracing algorithm. For simplicity, we assume that the light of each source follows a S$\rm \acute{e}$rsic profile, with a random ellipticity $q$ and position angle $\theta$. The size and S$\rm \acute{e}$sic index of the source are given based on their evolution with redshift \citep{mowla19}. Therefore, the final range of semi-major axis is from 0.3 to 0.8 arcsec, and the range of S$\rm \acute{e}$sic index varies from 1.0 for sources at higher redshifts to 4.5 for lower redshift sources. The superimposed surface brightness distribution ${S}_{\rm obs}$ within the field-of-view, summed from each source, then becomes the observable in our pixelated modeling method.

We have constructed the cluster strong lens system in a comprehensive manner, taking into account all the image families expected to be observed in next-generation surveys. Nonetheless, in the context of comparing traditional lens modeling and pixelated modeling, which can be computationally expensive, we randomly select only three of these image families as constraints for the following analyses. The redshifts of these selected sources are marked in Table~\ref{table1}.

% noise brammer12, burke19
% End of Sect 2.

\section{Methodology} \label{section3}

In this section, we describe the methods of analysing the simulations generated in Section~\ref{section2}. Firstly, we introduce the concept of measuring the cluster mass profile and magnification map based on the lensing formalism. We also discuss the capability of cluster strong lenses as a probe for constraining cosmological parameters. Subsequently, we provide a comprehensive explanation of our modeling methods, encompassing both the traditional lens modeling and a novel pixelated algorithm proposed in this work.

\subsection{Lensing theory}
The lens equation relates the position of source $\beta$ at redshift $z_{\rm s}$ to the position of image $\theta$ at redshift $z_{l}$:
\begin{equation} \label{eq3}
    \vec{\beta} = \vec{\theta} - \vec{\alpha}(\vec{\theta}),
\end{equation} 
where $\vec{\alpha}$ is the reduced deflection angle as a result of lensing potential. 
The mapping from unlensed coordinates on the source plane to that on the image plane can be described by the Jacobian matrix, 
\begin{equation}
    A \equiv \frac{\partial \vec{\beta}}{\partial \vec{\theta}} = (\delta_{ij} - \frac{\partial \alpha_{i}(\vec{\theta})}{\partial \theta_{j}}) = (\delta_{ij} - \frac{\partial ^ 2 \Psi (\vec \theta)}{\partial \theta_{i} \partial_{j}}), 
\end{equation}
defining the convergence $\kappa = \frac{1}{2} \Delta \Psi = \frac{1}{2}(\Psi_{11} + \Psi_{22})$ as a function of the effective lensing potential $\Psi$, to delineate the magnification of images, the shear vectors $\gamma_1 = \frac{1}{2}(\Psi_{11} - \Psi_{22})$ and $\gamma_2 = \Psi_{12} = \Psi_{21}$, where the complex shear $\gamma = \gamma_{1} + i\gamma_{2}$ describes the stretching of images, we can write the Jacobian matrix as 

\begin{equation}
A = \begin{pmatrix}
1 - \kappa - \gamma_{1} & - \gamma_{2}\\\\
    -\gamma_{2} & 1 - \kappa + \gamma_{1}
\end{pmatrix}.
\end{equation}
The magnification of the lens is then given by the inverse of the determinant of the Jacobian matrix,
\begin{equation}\label{eq6}
    \mu \equiv {\rm det} M = \frac{1}{{\rm det} A} = \frac{1}{(1-\kappa)^2 - \gamma^2}, 
\end{equation}
tangential and radial magnification factors are defined as
\begin{equation}
    \rm \mu_{t} = \frac{1}{\lambda_t} = \frac{1}{1-\kappa - \gamma}
\end{equation}
\begin{equation}
    \rm \mu_{r} = \frac{1}{\lambda_r} = \frac{1}{1-\kappa + \gamma},
\end{equation}
respectively. Regions with $\lambda_{\rm t} = 0$ and $\lambda_{\rm r} = 0$ are defined as critical curves in the lens plane, and caustics in the corresponding source plane, in which the magnifications are ideally infinite. These areas are particularly advantageous for observing distant and faint galaxies.

As the definition of convergence $\kappa = \frac{\Sigma}{\Sigma_{\rm crit}}$,
where $\Sigma_{\rm crit} = \frac{c^2}{4\pi G} \frac{D_{OS}}{D_{LS}D_{OL}}$ is the critical surface mass density of the Universe,  one can also calculate the cluster mass by summing the convergence  in an aperture.

\begin{figure}
	% To include a figure from a file named example.*
	% Allowable file formats are eps or ps if compiling using latex
	% or pdf, png, jpg if compiling using pdflatex
	\includegraphics[width=0.49\textwidth]{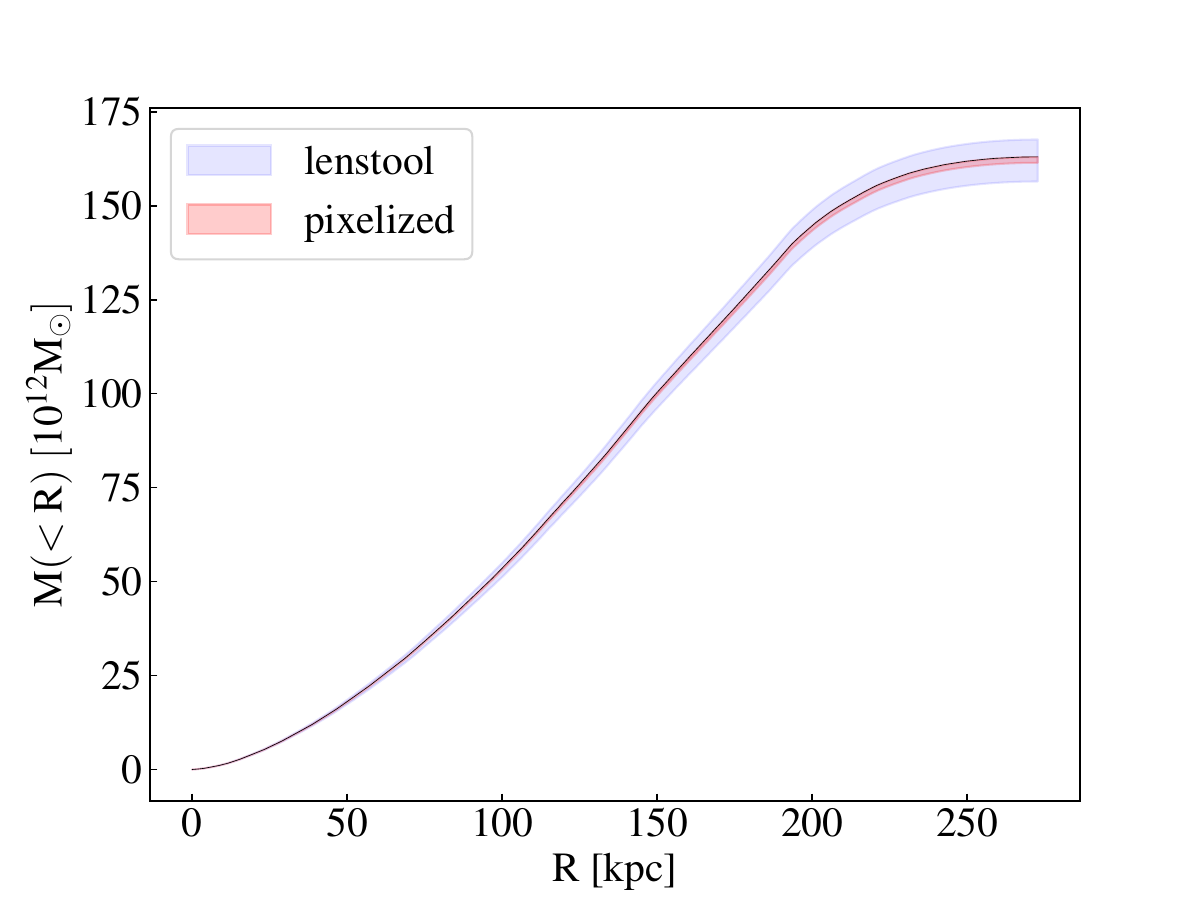}
    \caption{Circularly averaged mass profiles of the cluster recovered with blue: \textsc{Lenstool} and red: pixelated modeling. Shaded regions are obtained from the randomly selected 2000 realizations in the MCMC outputs. The input mass profile is indicated as the black line.}
    \label{fig4}
\end{figure}

\begin{figure*}
	% To include a figure from a file named example.*
	% Allowable file formats are eps or ps if compiling using latex
	% or pdf, png, jpg if compiling using pdflatex
	\includegraphics[width=0.98\textwidth]{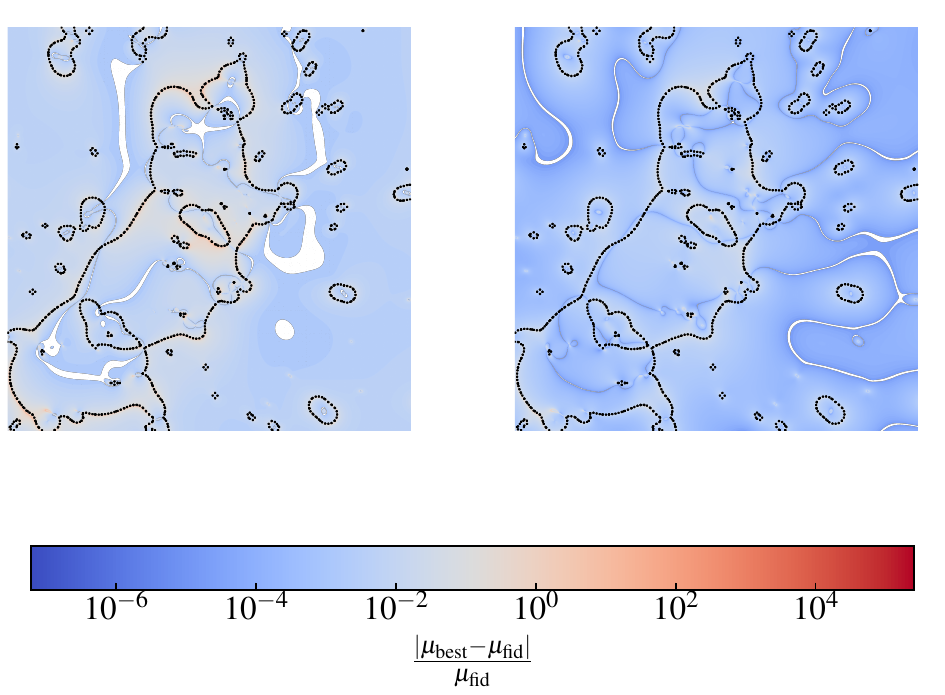}
    \caption{Log-scale difference between the recovered and fiducial magnification maps at $z_{\rm s} = 10$ with \textsc{Lenstool} (left panel) and pixelated modeling (right panel), with critical curves overlaid as black dots. Note: Regions with little difference ($<10^{-7}$) are plotted as white areas since they become Not-a-Number (NaN) in log scale.}
    \label{fig5}
\end{figure*}

\subsection{Strong lensing cosmological constraints}

We further write down the reduced deflection angle in the lens equation Eq.~\ref{eq3} as 
\begin{equation}\label{eq9}
    \vec{\alpha}(\vec{\theta}) = \frac{2}{c^2}\frac{D_{LS}}{D_{OL}D_{OS}} \nabla \phi(\vec{\theta}),
\end{equation}
wherein $c$ is the speed of light, and $\phi$ denotes the projected Newtonian lens potential. The angular positions of multiple images on the source plane and lens plane are interconnected through the lens potential and cosmic distances $D_{OL}$, $D_{LS}$ and $D_{OS}$. The angular diameter distance between redshift $z_{1}$ and $z_{2}$ in a flat Friedmann–Robertson–Walker cosmology is given by:
\begin{equation}
    D(z_{1},z_{2}) = \frac{c{\rm H}_{0}^{-1}}{1+z_{2}}\int_{z_{1}}^{z_{2}}dz [\Omega_{\rm m}(1+z)^{3}+\Omega_{\rm X}(1+z)^{3(w_{\rm X}+1)}]^{-1/2},
\end{equation}
where ${\rm H}_{0} = 100\ h\ \rm km\ s^{-1}$ is the Hubble constant at present day, $\Omega_{\rm m}$ and $\Omega_{\rm X}$ are the corresponding densities of total matter and dark energy, and $w_{\rm X}$ is the parameter for the equation of state of dark energy. 

The above equations demonstrate the capacity of cluster strong lenses to constrain the cosmological parameters $\lbrace \Omega_{\rm m}, w_{\rm X} \rbrace$, although the dependence of cosmology is entangled with the cluster mass. In order to disentangle the degeneracy between cluster mass distribution and cosmology, more than two families of multiple images have to be provided, via the family ratio $\Xi$:
\begin{equation}
    \Xi (z_{L},z_{S_{1}},z_{S_{2}};\Omega_{\rm m},w_{\rm X}) = \frac{D_{LS_{1}}}{D_{OS_{1}}}\frac{D_{OS_{2}}}{D_{LS_{2}}},
\end{equation}
where $z_{L}$, $z_{S_{1}}$, and $z_{S_{2}}$ are the lens redshift and two distinct source redshifts, $D$ represents the corresponding angular diameter distances. Alternatively, for the strong lensing system with only one family of multiple images, as proposed in \cite{golse02}, an additional prior on the mass distribution obtained from observations independent of strong lensing should be introduced to break the degeneracy between cluster mass and cosmology.

\begin{figure*}
	% To include a figure from a file named example.*
	% Allowable file formats are eps or ps if compiling using latex
	% or pdf, png, jpg if compiling using pdflatex
	\includegraphics[width=0.98\textwidth]{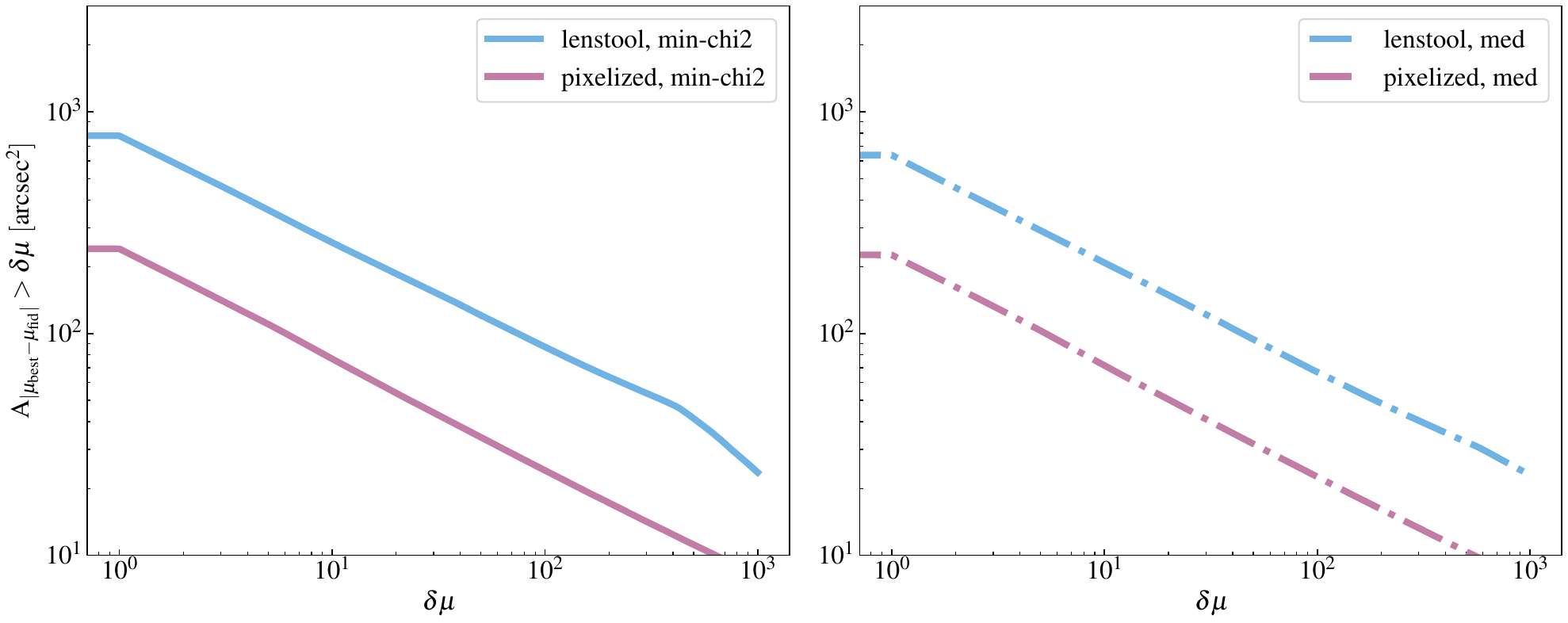}
    \caption{Total area as a function of the difference between the recovered and fiducial magnification maps. Left: comparison between \textsc{Lenstool} (blue) and pixelated modeling (pink) with the best-fit result defined as model with minimum $\chi^2$; right: same with left but with the best-fit result defined as medians of the parameters. }
    \label{fig6}
\end{figure*}

\subsection{Lens modeling}

In the context of providing predictions for the impact of point-like multiple images on future surveys, we exclude other systematic errors, and therefore adopt the same form of mass distribution as that used in our simulated clusters during the modeling process. To perform lens modeling, we utilize the Bayesian Markov Chain Monte Carlo (MCMC) sampler, which is implemented in the publicly available \textsc{Lenstool} software \citep{kneib96, jullo07, jullo09}. This modeling package allows us to sample the parameters related to both the cluster mass distribution and cosmology. 
As proposed in \cite{kneib11}, optimizing the model on the source plane can introduce biases due to the magnification factor, which tends to favor flat density profiles and large ellipticities. Hence, to improve parameter accuracy, we choose to perform the optimization on the image plane throughout the work, even though it requires more computational time since it involves an extra step of relensing, and models that generate different image configurations will be rejected \citep{kneib11}. 
The calculation of $\chi_{i}^{2}$ for a single multiple image family $i$ on the image plane is defined as follows:
\begin{equation}\label{eq12}
    \chi_{i} ^{2} =\sum_{j=1}^{{\rm n}_{i}} \frac{[\vec{\theta} _{\rm obs}^{j}-\vec{\theta}^{j}({\rm p})] ^{2} }{\sigma_{ij}^{2}}, 
\end{equation} 
where the position of the observed image $j$ $\vec{\theta} _{\rm obs}^{j}$ is a combination of the position of the simulated image with a random positional error of $\rm 0.1^{\prime \prime}$,
$\vec{\theta}^{j}({\rm p})$ is the position of image $j$ predicted by the model with parameter set \{p\}, $\sigma_{ij}$ denotes the positional error of this image, and the overall $\chi^2$ is obtained by summing over all image systems. 
We assume a smaller global Gaussian error of $\sigma_{ij} = 0.1^{\prime \prime}$ for the position of each multiple image in space-based measurements \citep{daloisio10} . Furthermore, we make the assumption that all available multiple images can be measured with spectroscopic redshifts, thus we do not treat the redshifts of multiple images as free parameters throughout the entire process.

\subsection{The pixelated modeling algorithm}

Given the significant computational complexity involved in modeling the observed multiple images pixel by pixel, current approaches often simplify the analysis by treating the multiple images as point sources. As depicted in Figure~\ref{fig2}, the likelihood is uniform in each direction under the assumption that the multiple image is treated as a point source. However, this assumption is only approximate due to the arc-like shapes produced by the lensing potential, resulting in nonuniform likelihood in each direction. As a result, defining the arc-like shape error solely based on a Gaussian function can lead to biased constraints.

The bias introduced by the point-like multiple image assumption can be buried under various unresolved systematic errors arising from the limitations of observations. Nonetheless, as we look forward to the surveys from the next generation telescopes, such as the Stage-IV surveys, strong lens modeling of galaxy clusters becomes increasingly critical for a deeper understanding of the Universe. In this context, it is imperative that we thoroughly comprehend and address all types of systematic errors. To mitigate such bias, this work introduces an alternative approach: pixelated lens modeling.

The steps in our method closely follow those of \textsc{Lenstool}, but with a key difference in the calculation of $\chi^2$. Instead of comparing the observed and model-predicted positions of multiple images, we compute the corresponding extended surface brightness of every image and sum them to create a total image. The comparison between models and observations is then made on a pixel-by-pixel basis,
\begin{equation}
    \chi ^{2} =\sum_{i=1}^{{\rm n_{pix}}} \frac{[{S} _{\rm obs}^{i}-{S}^{i}({\rm p})] ^{2} }{\sigma_{i}^{2}}, 
\end{equation}
where $S_{\rm obs}^{i}$ is the observed surface brightness on pixel $i$ of the total image, $S^{i}({\rm p})$ is the model-predicted surface brightness on pixel $i$ with model parameters ${\rm p}$, and $\sigma_{i}$ is the error on pixel $i$. Similar to the implementation in \textsc{Lenstool}, the observed surface brightness is modeled as the sum of the ideal brightness distribution and a noise field. The noise map is tailored for a JWST setup, incorporating a readout noise of 15.77 $e^-$, a zero-point magnitude of 28.0, and a total exposure time of 7, 537 seconds across nine exposures. This setup is designed to simulate the noise contributions from both the sky and read noise within the aperture, while we assume that the non-linear effects, e.g., charge transfer efficiency, hot pixels, residuals in flat fields, are mitigated before the modeling process. The calculation of $\chi^2$ is performed by integrating over all the pixels in the entire field of $72^{\prime \prime} \times 72 ^{\prime \prime}$, with a pixel size of $0.18^{\prime \prime}$. We note that this resolution is currently not at the JWST level, which is considered due to computational costs. As introduced in Section~\ref{section22}, to reduce complexity, we assume that the light distribution of each source follows a same S$\rm \acute{e}$rsic profile, and we keep these parameters fixed while modeling. The Probability Distribution Function (PDF) of the lens parameters is optimized using the Monte-Carlo-Markov-Chain (MCMC) algorithm implemented in the \textsc{emcee} \citep{emcee} package.

% End of Sect 3.

\begin{figure*}
	% To include a figure from a file named example.*
	% Allowable file formats are eps or ps if compiling using latex
	% or pdf, png, jpg if compiling using pdflatex
	\includegraphics[width=0.98\textwidth]{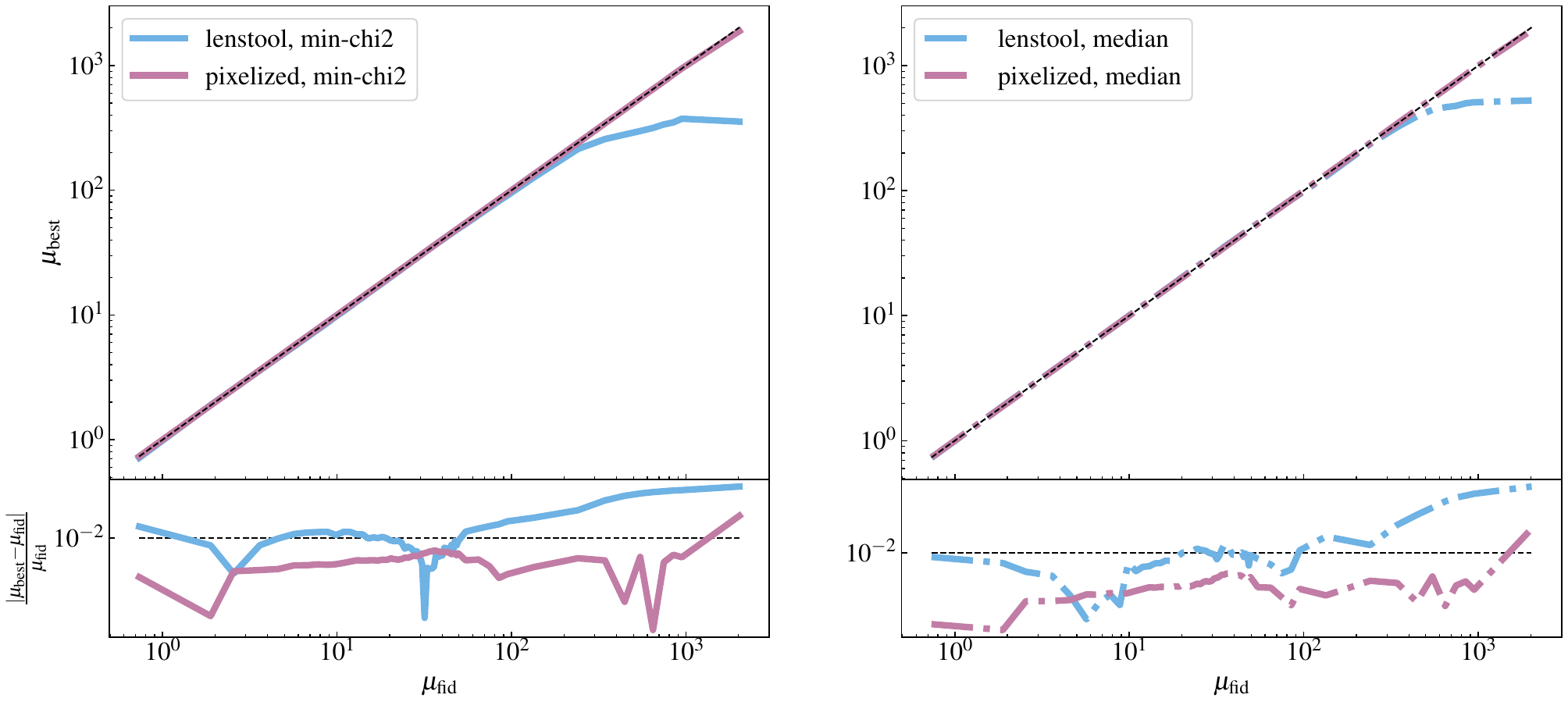}
    \caption{Conditional distribution \textit{P }($\mu_{\rm best}$|$\mu_{\rm fid}$) (top) with the error of best-fit recovered magnification map (bottom), where the corresponding values of magnification on pixels with $\mu_{\rm fid}$ are calculated as $\mu_{\rm best}$ in the recovered magnification map. The panels and color coding are the same as in Figure~\ref{fig6}.}
    \label{fig7}
\end{figure*}

\section{Results} \label{section4}

Before analyzing the impact of assuming multiple images as point sources on the recovery of cluster mass profiles, magnification maps, and cosmography, we first visualize the parameter PDFs of the simulated cluster (Abell 2744-like) from \textsc{Lenstool} (blue) and pixelated modeling (pink) in Figure~\ref{fig3}. To ensure a comprehensive comparison, we also include the modeling results from \textsc{Lenstool} with \textsc{emcee} as the sampler, referred to as "lenstool-emcee" (green). This additional analysis aims to address the possibility that the investigated bias is induced by the choice of sampling algorithm.  Note that the cluster model is simplified throughout the entire work. Specifically, we vary the parameters of the first main halo and cosmology while keeping all other parameters fixed. Additionally, the number of image families used as constraints is reduced to three. These reductions in the number of constraints and free parameters during optimization are intended to ensure computational efficiency. In our analysis, we generate the posteriors by running 100 walkers for 10000 steps, while discarding the first 3000 steps as part of the "burn-in" phase.

Our analysis reveals that parameters obtained with \textsc{Lenstool} and lenstool-emcee exhibit similar degeneracies, the constraint on the main halo's truncation radius $\rm r_{cut}$ with lenstool-emcee is even less accurate compared to that with \textsc{Lenstool}. Therefore, it is reasonable to exclude the possibility that the observed bias is solely due to the sampling algorithm. On the other hand, we find a significant improvement in the posteriors obtained through our pixelated modeling approach. The distribution of each parameter from \textsc{Lenstool} is deviated from a Gaussian distribution centred on the input value, as indicated by the grey dashed line. In contrast, such a deviation is not apparent in the parameters derived from pixelated modeling, although we notice a slight deviation from a Gaussian distribution for the constraint on $\rm r_{cut}$ at large values. This can be understood as constraints from strong lensing primarily lying in the central region of a cluster field. In addition, the parameter distributions are narrower when comparing pixelated modeling to traditional modeling. 
This is attributed to the fact that pixelated modeling employs information from every pixel in the image, as opposed to relying on a handful of multiple image positions. However, cautions should be taken that since we are in the ideal case, discussing the accuracy of lens modeling under the assumption of point-like multiple images, it is worth noting that this analysis does not take into account any sources of systematic errors from the observational side. Therefore, the exact level of precision can only be determined by conducting more realistic simulations.

\subsection{Mass profiles}
Investigating the mass distribution of strong lensing clusters is fundamental for understanding the nature of dark matter and the formation and evolution of  galaxy clusters \citep{okabe20, newman13}. As a way to compare the results from traditional and pixelated modeling methods, we present the recovered mass profiles in Figure~\ref{fig4}. In this figure, we plot the enclosed mass obtained using the two modeling methods (blue shaded region: \textsc{Lenstool}, red: pixelated modeling) as a function of its radius R, where the input mass distribution is compared with the masses recovered with 2000 randomly selected realizations. 

We find that both methods can provide accurate constraints on the cluster mass profile, given that the same form of cluster mass model with the simulation is used for fitting. This suggests that the investigated bias has little impact on the recovery of a strong lensing cluster mass profile. We also observe that the results from pixelated modeling are more precise compared to traditional modeling. This is expected because pixelated modeling incorporates more information by using the intensity of each pixel as a constraint, allowing for a larger degree of freedom.

\subsection{Magnification maps}

The magnification of strong lensing clusters has become a primary scientific objective in projects such as JWST \citep{stek23, bradac23, bradley23}, HFF \citep{lotz17, bouwens22, yang22, fox22}, and RELICS \citep{salmon20, coe19, neufeld22}, providing valuable opportunities for observing and studying high-redshift objects. Therefore, it is essential to investigate the robustness of the recovered magnification maps.

Before comparing the results between the traditional method, i.e., \textsc{Lenstool}, and the pixelated method, we  need to establish a definition for the best-fit lens model based on the output from MCMC sampling. In some literature \citep{acebron19, gonzalez21}, it is defined as the model parameters that maximize the probability distribution function (PDF),  often referred to as the Maximum A Posteriori (MAP) estimate. In \textsc{Lenstool}, this is represented as the parameter set with minimum $\chi^2$. However, the MAP estimate may not always be the best summary of the posterior estimation. In some cases, the best-fit model is defined using the mean, median, or mode of the samples \citep{caminha16, raney20}. In this work, we present the results with both the best-fit model designated as the one with the minimum $\chi^2$, and as the medians of the samples. The magnification maps illustrated in this subsection are obtained using the best-fit models based on Eq.~\ref{eq6}, where we assume a source redshift of $z_{\rm s} = 10$. This redshift is chosen as it represents a typical value for discovering and studying high-redshift lensed objects. We utilize the absolute value of magnification throughout the analysis.

Figure~\ref{fig5} provides an illustration of the difference between the best-fit (medians) and the fiducial magnification maps obtained with \textsc{Lenstool} (left) and pixelated modeling (right) at the core of the cluster field, where critical curves are also superimposed as black dots for reference. To enhance clarity, the difference is normalized by the fiducial magnification map. In most regions of the two panels, the magnification difference is below $10^{-2}$. However, it is worth noting that in regions close to critical curves in the left panel, the difference can exceed $10^4$. These regions near critical curves are of particular interest as they exhibit extreme magnification (up to thousands), that are expected to host individual high-redshift stars \citep{meena23, kelly22}, or faint galaxies from the epoch of reionization (EoR) \citep{yan23, yue14}, offering valuable insights into various outstanding questions such as the galaxy formation. Even in the cluster core regions that are not close to the critical curves, there are also large areas with a magnification difference $|\mu_{\rm best} - \mu_{\rm fid}| / \mu_{\rm fid} > 10^2$. Therefore, the substantial disparity observed between the recovered and fiducial magnification maps raises concerns about the potential misinterpretation of the properties of lensed objects in these critical regions. On the other hand, we notice that pixelated modeling offers improved accuracy in predicting magnification maps, as shown in the right panel, especially in regions near critical curves. This helps reduce the likelihood of extreme errors that are more prevalent in traditional modeling using \textsc{Lenstool}. 

We calculate the total area representing different magnification differences, denoted as $\delta \mu = |\mu_{\rm best} - \mu_{\rm fid}|$, within the four cases: using \textsc{Lenstool} or pixelated modeling as the fitting algorithm, and preferring minimum $\chi^2$ or medians as the best fit. These results are shown in Figure~\ref{fig6}. For both criteria of best-fit, we find that \textsc{Lenstool} fails to produce magnification maps that are as accurate as those obtained by pixelated modeling.

We proceed with a detailed pixel-by-pixel comparison of the magnification maps. To generate the conditional distribution P($\rm \mu_{X}|\mu_{ref}$) shown in Figure~\ref{fig7}, we select all the pixels with a specific magnification value $\mu_{\rm fid}$ in the fiducial map (e.g., $\mu_{\rm fid} = 1$) and calculate the median value of these corresponding pixels $\mu_{\rm best}$ in the recovered map. This process is repeated for magnification values spanning the range from $\mu_{\rm fid} = 0$ to $\mu_{\rm fid}$ > 1000. Consistent with the observations from Figure~\ref{fig5} and \ref{fig6}, we find that pixels with higher magnification, particularly those near critical curves, exhibit inaccuracies regardless of whether the minimum $\chi^2$ or median values are used as the best-fit model. These discrepancies suggest that the properties of lensed sources identified in these regions should be regarded with caution, as they may not be reliable. The biases in the \textsc{Lenstool}  recovered maps become more pronounced as the magnification $\mu$ increases. At $\mu=10$, the biases are $\rm |\mu_{best} - \mu_{fid}| / \mu_{fid} = 0.005$ (\textsc{Lenstool}) and $\rm |\mu_{best} - \mu_{fid}| / \mu_{fid} = 0.0007$ (pixelated modeling), while the biases reach $\rm |\mu_{best} - \mu_{fid}| / \mu_{fid} = 0.4624$ (\textsc{Lenstool}) and $\rm |\mu_{best} - \mu_{fid}| / \mu_{fid} = 0.0009$ (pixelated modeling) at $\mu = 1000$, indicating that only with pixelated modeling can we obtain a magnification with accuracy.

\begin{figure*}
	% To include a figure from a file named example.*
	% Allowable file formats are eps or ps if compiling using latex
	% or pdf, png, jpg if compiling using pdflatex
	\includegraphics[width=0.98\textwidth]{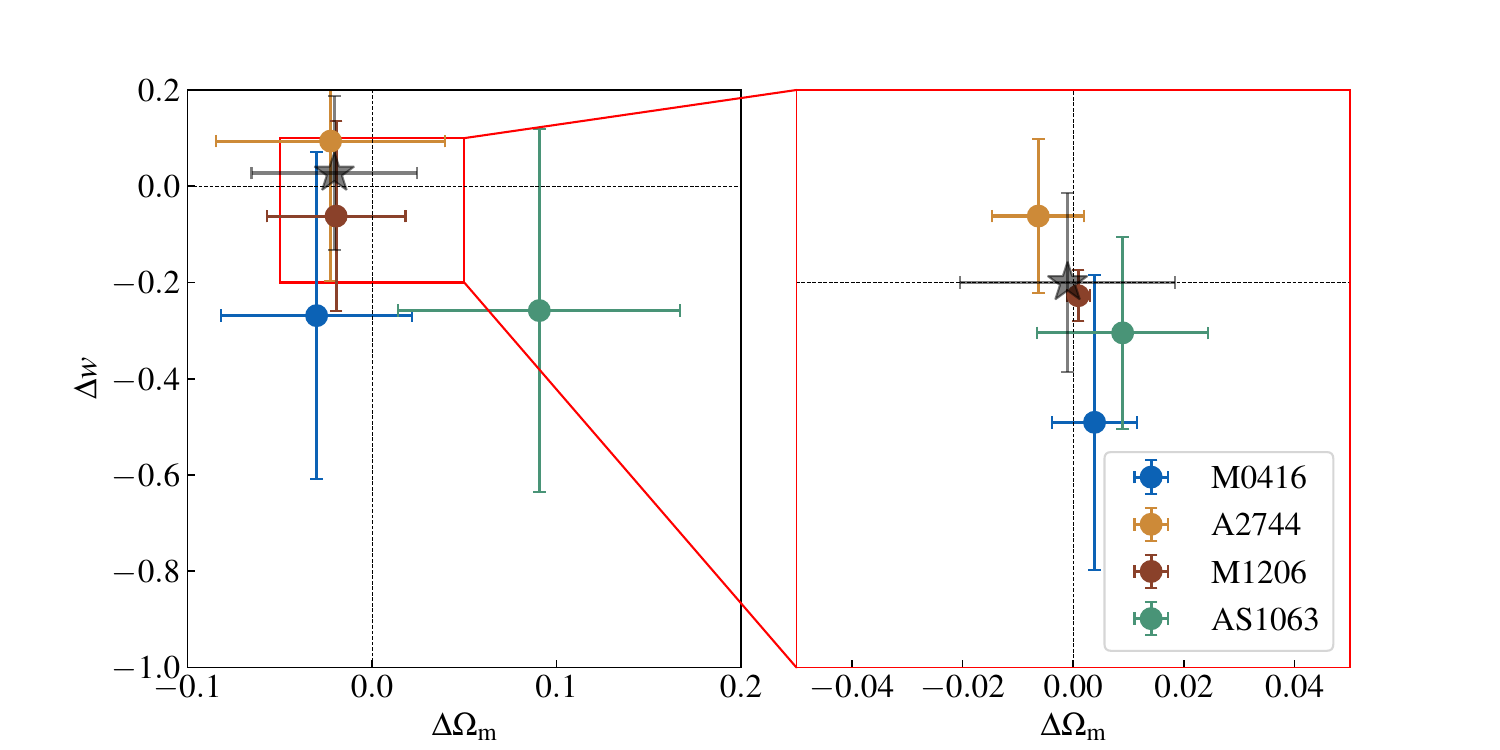}
    \caption{Cosmological constraints derived from the four simulated clusters (Left: obtained with \textsc{Lenstool}, right: obtained with the pixelated modeling). Each colored point denotes the difference between the median values with 1-$\sigma$ errors of the cosmological parameters $\Omega_{\rm m}$ and $w$ from an individual cluster system and the fiducial values, the input cosmology is marked with the black dashed lines. The grey star represents the result from the stack of the four clusters. }
    \label{fig8}
\end{figure*}

\subsection{Cosmological constraints}

Since it was initially proposed by \cite{link1998}, cluster strong lensing has been extensively studied to demonstrate its feasibility on cosmography, as well as to explore related statistical and systematic errors \citep{gilmore09, daloisio11, acebron17, magana18}. In this subsection, we analyze the cosmological constraints from our simulated lens systems, and investigate the impact of the assumption of point-like multiple images on this issue.

We present the cosmological constraints of the four clusters in the simulation, obtained from both modeling algorithms, in Figure~\ref{fig8}. To derive the PDFs P$_{i}$(cosmo) for the cosmological parameters from each cluster $i$, we marginalize over the mass parameters within the corresponding MCMC chains. We observe that despite the absence of systematic errors in the modeling process, the posterior from each individual system is not perfectly aligned with the fiducial cosmology (marked as the black dashed lines). The median values of $\Omega_{\rm m}$ are more likely to be larger than the fiducial $\rm \Omega_{m, fid} = 0.3$, while the median values of $w$ lie in the region of lower values compared to the fiducial $w_{\rm fid} = -1.0$.

Given the complexity involved in the process of cluster strong lens modeling, a stacked cosmological analysis has been proposed to enhance the figure of merit as well as mitigate potential systematic errors among different clusters. Therefore, we also incorporate the combining constraint from the four clusters. The combined PDF is derived by taking the product of PDFs from all the four systems,
\begin{equation}
	\rm P_{total}(cosmo) = \prod \limits_{i=1}^N P_{\it i}(cosmo).
\end{equation}
In the left panel obtained with \textsc{Lenstool}, the cosmological constraint arising from a stack of four HFF-like cluster lens systems ($\Delta \Omega_{\rm m} = 0.04$, $\Delta w = 0.16$ in the ideal case) demonstrates its competitiveness with other available cosmological probes (CMB, $\Delta \Omega_{\rm m} = 0.04$, $\Delta w = 0.27$, \cite{planck18}). Combining the systems seems to help eliminate the deviation from the fiducial values within errors, we obtain the cosmological constraints $\Omega_{\rm m} = 0.28 \pm{0.04}$ and $w = -0.97 \pm{0.16}$, exhibiting slight deviations but are consistent with the fiducial cosmology within the 1-$\sigma$ confidence level. However, we argue that this biased cosmological constraint is an inevitable issue when the Gaussian light profile assumption is used in modeling methods for multiple images. It is important to recognize that the direction of deviation from the truth in each lens system is not linear. Therefore, the cosmological constraints by combining several systems can be even more biased according to the selection bias of strong lensing clusters.

In turn, the parameter PDFs of each lens system follow a more typical Gaussian distribution, with medians that closely align with the input values when using pixelated modeling. This correspondence can be attributed to the comprehensive consideration of the intensity of every pixel in the pixelated modeling approach, eliminating the need for the Gaussian light profile approximation for multiple images. By removing this source of systematic error, we can obtain cosmological constraints with median values of $\Omega_{\rm m} = 0.30$ and $w = -1.0$ that are expected to be unbiased.

Based on the above analyses, we emphasize the importance of developing lens modeling algorithms that operate on pixel scale. Such algorithms have the potential to alleviate the deviation from the truth, and provide more accurate and reliable cosmological constraints. By incorporating more information, specifically treating each pixel as a constraint, we can obtain more stringent cosmological contours compared to traditional algorithms that rely solely on the positions of images. Consequently, even with a limited number of observed clusters, we can provide competitive constraints through cluster strong lensing.

% End of Sect 4.

\section{Discussion and Conclusion} \label{section5}

With the designed large field of view and deep imaging capabilities, stage-IV surveys, such as Euclid, CSST, and JWST, will have the ability to observe numerous multiple images and giant arcs in cluster fields. This presents a valuable opportunity to gain deeper insights into systematic errors from the observational side and holds great promise in the cluster strong lensing field. In this context, the assumption of multiple images to be point sources and approximating their light profiles as a two-dimensional Gaussian distribution in traditional modeling, buried in other systematic errors, will emerge as a non-negligible issue. In this work, we employ a series of simulated cluster strong lenses to assess the impact of the above bias and propose a possible solution to address it. 

In the first part, we take advantage of the cluster lens system similar to Abell 2744 at redshift $z=0.4$ to investigate the mass map obtained from lens modeling, as illustrated in Figure~\ref{fig4}. We find that the cluster mass profile can be accurately recovered using \textsc{Lenstool}, implying that the bias in discussion has little influence. Regarding the recovery of the magnification map, as illustrated in Figure~\ref{fig5} to Figure~\ref{fig7}, we observe that the magnification map of the targeted cluster can be measured accurately in most regions of the strong lensing field, if valuable constraints from the data are provided. However, in regions close to critical curves, we find that the magnification difference is still too large to robustly study the properties of high-redshift objects. To address this issue, we exploit a pixelated approach to model the lens system, which takes into account the light distribution on each pixel of the image. We compare the performance of the pixelated approach to traditional lens modeling and demonstrate the improvement achieved with the new approach. Next, we focus on the cosmological constraints obtained from cluster strong lenses. As shown in Figure~\ref{fig8}, we find that strong lensing clusters can provide powerful constraints compared to other tools like Cosmic Microwave Background \citep[CMB][]{planck18, hinshaw13, komatsu11}, Baryonic Acoustic Oscillation \citep[BAO][]{alam17, abbott21}, weak gravitational lensing \citep{destroxel18, joudaki18, bocquet19}, and others. Nonetheless, the bias is also introduced into the cosmological constraint thanks to the Gaussian-like approximation of multiple images. 
The pixelated approach also demonstrates enhancement in the same figure, where the peak of parameter PDF align with the true value. Moreover, since the pixelated approach utilizes more information, we can expect competing cosmological constraints even with a single cluster lens system.

Accurate lens modeling requires careful treatments. Our results have demonstrated that the Gaussian-approximated likelihood employed in traditional modeling methods introduces bias to the modeling results, particularly in the recovered magnification maps and cosmological constraints. Furthermore, additional caution should be exercised during the modeling process. For instance, the proper way for model optimization is completed on the image plane, which compares each observed multiple image to its corresponding model-predicted one iteratively and is quite time-consuming. In order to improve efficiency, some studies \citep{jullo10, gilmore09, acebron17} have opted to perform optimization on the source plane by minimizing the discrepancy among model-predicted source positions without the need for an additional step of re-lensing. However, this alternative approach can lead to a biased optimization, favoring mass models with flat density profiles and large ellipticities \citep{kneib11}. Based on these considerations, we maintain our approach of fitting the clusters on the image plane throughout the paper, to ensure the attainment of the most accurate results. 

The accurate modeling of strong lensing clusters has always posed a challenging task due to the complex structures of the clusters and the large number of strongly lensed multiple images that serve as constraints. In this paper, we propose the idea of creating a new modeling approach to solve the bias induced in traditional modeling. The approach is coded on the basis of \textsc{Lenstool}, but instead of computing the $\chi^2$ and corresponding likelihood function using the positions of multiple images, we prefer to fit the image plane pixel by pixel. Even with the currently available algorithms, the process of fully exploring the parameter space and obtaining results from MCMC or other sampling methods can be time-consuming, often taking days to weeks to fit on the image plane. In our work, modeling the simulated A2744-like cluster with \textsc{Lenstool} requires 9 hours and 48 minutes, whereas the same system with the proposed pixelated method takes 25 hours and 34 minutes. Furthermore, considering all image families would result in a modeling time of months for a cluster lens. Hence, any opportunity to accelerate this process is necessary. To address this challenge, one potential solution is to upgrade the algorithm using GPU-based acceleration. Promising tools for this upgrade include JAX \citep{jax2018github} and Numpyro \citep{bingham19}, which can expedite both the processes of computing the deflection field and the sampling. These novel tools have the potential to significantly enhance speed and have already been applied to multiple purposes in astrophysics \citep{pasha23, campagne23, numpyro}. 

% Furthermore, our current work primarily relies on analytical simulations. While these simulations serve as a valuable tool for understanding the investigated bias and testing our proposed modeling method, they may not fully encapsulate the intricacies of realistic lensing clusters. For example, secondary masses along the line of sight of the cluster can contribute to the lensing signal, and impact the detectability of strong lensing features \citep{li19, daloisio10, bayliss14}, therefore need to be properly modeled to avoid misinterpretations. Also, the dense environments of galaxy clusters result in a wide variety of cluster member galaxies \citep{kuchner17}, thus using a unified set of scaling relations, as described in Eq.~\ref{eq2}, to construct the galaxies is not sufficient. In future work, we plan to extend our simulations to account for these complexities. However, it is worth noting that using parametric modeling methods on these more complex clusters could introduce systematic biases. To improve modeling results, further considerations, for instance, a combination of parametric and free-form approaches \citep{benjamin21}, or applying local potential correction \citep{vegetti09, galan22}, should be included into the modeling.
Our current work primarily relies on analytical simulations to comprehend the bias due to point-like image approximation and test the feasibility of our proposed modeling approach. However, it must be noted that these simulations may not fully capture the intricate details of real clusters in the Universe. Factors such as structures along the line of sight and the surrounding environment can influence the characteristics of strong lensing signals \citep{li19, daloisio10, bayliss14, kuchner17}, requiring proper modeling. Additionally, adopting parametric models only might underfit a cluster-scale strong lensing system with the structures mentioned above, so it is imperative to consider a combination of parametric and free-form approaches \citep{benjamin21}, or incorporate local potential correction methods \citep{vegetti09, galan22}. Our future research endeavors will be dedicated to addressing the aforestated issues.

\section*{Acknowledgments}
The authors thank the anonymous referee who provided useful suggestions that improved this manuscript. We acknowledge the support by National Key R\&D Program of China No. 2022YFF0503403 and the Ministry of Science and Technology of China (grant Nos. 2020SKA0110100). This work is supported by China Manned Space Project (No. CMS-CSST-2021-A01, CMS-CSST-2021-A04, CMS-CSST-2021-A07, CMS-CSST-2023-A03). HYS acknowledges the support from NSFC of China under grant 11973070, Key Research Program of Frontier Sciences, CAS, Grant No. ZDBS-LY-7013 and Program of Shanghai Academic/Technology Research Leader. RL acknowledges the support of National Nature Science Foundation of China (Nos 11988101,12022306), CAS Project for Young Scientists in Basic Research (No. YSBR-062). AA has received funding from the European Union’s Horizon 2020 research and innovation programme under the Marie Skłodowska-Curie grant agreement No 101024195 — ROSEAU. EJ acknowledges the support of the Centre National d'Etude Spatiale (CNES).

\section*{Data Availability}
The data underlying this article will be shared on a reasonable request to the authors.

%%%%%%%%%%%%%%%%%%%%%%%%%%%%%%%%%%%%%%%%%%%%%%%%%%

%%%%%%%%%%%%%%%%%%%% REFERENCES %%%%%%%%%%%%%%%%%%

% The best way to enter references is to use BibTeX:

%\bibliographystyle{mnras}
%\bibliography{example} % if your bibtex file is called example.bib

% Alternatively you could enter them by hand, like this:
% This method is tedious and prone to error if you have lots of references

%%%%%%%%%%%%%%%%%%%%%%%%%%%%%%%%%%%%%%%%%%%%%%%%%%

%%%%%%%%%%%%%%%%% APPENDICES %%%%%%%%%%%%%%%%%%%%%

%%%%%%%%%%%%%%%%%%%%%%%%%%%%%%%%%%%%%%%%%%%%%%%%%%

% Don't change these lines
\bibliography{main}
\bsp	% typesetting comment
\label{lastpage}
\end{document}